\documentclass[spanish]{maciarticle}

\usepackage{amsmath,amsbsy,amscd,amssymb,graphicx,epsfig,color}
\usepackage{soul} 

\newcommand{\bb}[1]{{{#1}}} 

\makeatletter

\begin{document}

\title{Modelando procesos cognitivos de la lectura natural con GPT-2}

\author[$1, 2 ,\dag$]{Bruno Bianchi}
\author[$1$]{Alfredo Umfurer}
\author[$1, 2$]{Juan Kamienkowski}

\affil[$\dag$]{\textcolor{blue}{bbianchi@dc.uba.ar, reading.liaa.dc.uba.ar}}

\affil[$1$]{Laboratorio de Inteligencia Artificial Aplicada, DC-ICC, UBA-CONICET, Buenos Aires, Argentina}

\affil[$2$]{Maestría en Explotación de Datos y Descubrimiento del Conocimiento, FCEN-UBA, Buenos Aires, Argentina}

\maketitle

\begin{abstract}
El avance del campo del Procesamiento del Lenguaje Natural ha permitido desarrollar modelos de lenguaje con una gran capacidad de generar texto. Desde hace algunos años, la Neurociencia utiliza estos modelos para comprender mejor los procesos cognitivos. En trabajos previos encontramos que modelos como Ngrams y redes LSTM son capaces de modelar parcialmente la Predictibilidad al ser utilizada como co-variable para explicar los movimientos oculares de lectores. En el presente trabajo profundizamos esta línea de investigación utilizando modelos basados en GPT2. Los resultados muestran que ésta arquitectura logra mejores resultados que sus predecesores.

\end{abstract}
\begin{mathsubclass}
68T50 - 92C20 
\end{mathsubclass}

{\thispagestyle{empty}} 

\section{Introducción}
El lenguaje es una de las características distintivas del ser humano. Esta capacidad con la que contamos nos permite comunicarnos de forma compleja con el fin de expresar pensamientos \cite{hauser2002faculty}. A lo largo de la historia diferentes campos científicos, como la lingüística, la psicología, y la neurociencia, han abordado el estudio del lenguaje desde diferentes ópticas. 

El campo del Procesamiento del Lenguaje Natural (PNL) ha evolucionado paulatinamente hacia algoritmos y modelos que son capaces de replicar el lenguaje humano con una gran fidelidad \cite{merityRegOpt2014,radford2018improving,radford2019language,brown2020language}. A pesar de que el objetivo general de ésta área no es comprender el lenguaje, sino simularlo computacionalmente, los modelos del estado del arte nos brindan una herramienta que puede resultar útil para comprender el cerebro. Estas herramientas se suman a los esfuerzos realizados desde la neurociencia cognitiva, la cual aborda el estudio del lenguaje con el objetivo general de comprender con detalle los mecanismos cerebrales que lo hacen posible. Y, en última instancia, generando un ciclo virtuoso en el que el conocimiento del lenguaje y los mecanismos cerebrales asociados aportan al desarrollo de algoritmos \cite{toneva2019interpreting}.

En este sentido, se ha estudiado cómo diferentes algoritmos del campo del PLN son capaces de estimar la Predictibilidad de una palabra. Esta variable se propone como un reflejo de cuán probable es para un lector saber una determinada palabra antes de leerla. Se estima mediante un experimento denominado \textit{cloze-task}, en el que se le pide a los participantes que completen sentencias incompletas con una única palabra \cite{taylor1953cloze}, y se calcula como la frecuencia con que se responde la palabra originalmente continuaba el texto (\textit{cloze-Pred}). Realizamos su modelado computacional (comp-Pred) mediante el uso de Modelos de Lenguaje Causales.

En las últimas décadas, diferentes investigaciones han modelado la \textit{cloze-Pred} mediante modelos computacionales simples. Hasta el momento todos estos intentos han logrado modelados parciales \cite{ong2008conditional,hofmann2017benchmarking,bianchi2020human,hofmann2021language,algan2021prediction,umfurer2021using}. En los últimos años ha aumentado el interés en realizar estos estudios en base a modelos de aprendizaje profundo. Hofmann y colaboradores \cite{hofmann2017benchmarking,hofmann2021language} entrenaron modelos de N-grams, Topic Models (LDA) y Redes Neuronales Recurrentes con textos de Wikipedia y subtítulos de películas. 
Al analizar la varianza de los tiempos de fijación explicada por estas probabilidades, concluyeron que los algoritmos computacionales pueden explicar mejor los movimientos oculares que la cloze-Pred. 

En estudios previos realizamos un análisis similar al de Hofmanm y colaboradores \cite{bianchi2020human,umfurer2021using}. Sin embargo, no analizamos directamente la varianza explicada por cada Predictibilidad (Cloze o Computacional), ya que este análisis puede pasar por alto que alguna de las variables utilizadas esté capturando porciones diferentes de la variabilidad de los datos de movimientos oculares. En este sentido hemos observado que, a diferencia de la cloze-Pred, todas las comp-Pred analizadas (modelos utilizados: N-grams, LSA, word2vec, FastText, AWD-LSTM) capturan parte de la varianza explicada originalmente por la Frecuencia Léxica de las palabras. Este efecto se ve levemente reducido al utilizar un modelo de lenguaje basado en redes LSTM entrenado con Wikipedia en español y reentrenado con textos de un dominio similar al de evaluación (textos narrativos). Nuestros trabajos también muestran que los residuos de los modelos lineales utilizados cuentan con varianza explicada por la cloze-Pred.

Actualmente nos encontramos en un momento en el que las arquitecturas basadas en \textit{transformers} han monopolizado el campo del PLN \cite{vaswani2017attention,radford2018improving,devlin2018bert}. El presente trabajo tiene como objetivo extender nuestros esfuerzos previos, en primer lugar, mediante la utilización de arquitecturas basadas en \textit{transformers}, y en segundo lugar
, en lo referente al corpus de entrenamiento. En particular, nos proponemos generar las comp-Pred con un modelo GPT2 entrenado en español, y reentrenado con dos corpus propios, uno del mismo dominio literario que los textos de evaluación y otro de la misma variante de español que los participantes (español rioplatense).

\section{Métodos}
\subsection{Datos: Textos, estimación de cloze-Pred y movimientos oculares}
En el presente trabajo utilizaremos los datos de cloze-Pred y Movimientos Ocualres publicados previamente por Bianchi y colaboradores \cite{bianchi2020human}. 
Se registraron los movimientos oculares (en particular la duración de las fijaciones) de 36 participantes durante la lectura de 8 textos narrativos. Con estos datos se calculó el Tiempo de Lectura en Primera Pasada (FPRT: suma de todas las fijaciones sobre una palabra en la primera pasada) para cada palabra leída por cada sujeto. Resultando en 54,121 elementos totales, $1503\pm618$ por sujeto, $6765\pm3226$ por texto, $20\pm35$ vistas por palabras, de $2588$ palabras únicas. Esta variable será utilizada como variable dependiente en los modelos estadísticos. Por otro lado, se tomaron las predicciones para cada palabra (\textit{cloze-Pred}) a través de un experimento online. Se obtuvieron un promedio de 13 respuestas para cada palabra (rango: 8–37). Este corpus cuenta con cada uno de los 8 textos completamente anotados con variables de interés: \bb{\textit{Distancia Sacada:} cantidad de caracteres recorridos por el ojo previo a la fijación actual; \textit{Longitud:} cantidad de caracteres de la palabra fijada; \textit{Frecuencia:} frecuencia léxica de la palabra en LexEsp; \textit{Rel pos:} posiciones relativas en la línea, el texto o la oración; \textit{Long:Frec:} interacción entre Longitud y Frecuencia; \textit{Pred:} predictibilidades Cloze y computacionales. Para más detalles, ver \cite{bianchi2020human}.}

\subsection{Predictibilidad Computacional}
En el presente trabajo se introdujo la arquitectura de modelo de lenguaje GPT2, y se la comparó con las anteriores. Se utilizó el modelo entrenado por el consorcio \textit{DeepEsp}, disponible en el repositorio de \textit{HuggingFace} \footnote{https://huggingface.co/DeepESP/gpt2-spanish}. Este modelo fue entrenado con 11.5GB de textos en español, entre Wikipedia (3.5GB) y libros (8GB) de diferentes dominios (narrativo, historias cortas, teatro, poesía, ensayos, y popularización).

Para el primer reentrenamiento de este modelo se utilizó el corpus de textos narrativos utilizado en trabajos previos \cite{bianchi2020human}. \bb{Al contar con un gran número de textos bajo licencia, éste corpus no es de acceso público.} El mismo cuenta con 2082 cuentos (600MB), de variada naturaleza, tanto en el género literario como en la nacionalidad de los autores. Adicionalmente, realizamos un segundo reentenamiento, independiente del primero (es decir, partiendo nuevamente del modelo original), con textos descargados de \textit{blogs} argentinos (28MB). \bb{Este último corpus se encuentra en preparación para ser abierto a acceso público. Ambos reentrenamientos fueron realizados sobre todos los parámetros del modelo original.}

\subsection{Modelos Lineales Mixtos}
En el presente trabajo tomamos como métrica los resultados de ajustar Modelos Lineales Mixtos sobre el logaritmo de la variable FPRT (\textit{lme4} v.3.1-144, R v.3.6.3 \cite{bates2014lme4}), obteniendo un t-valor por co-variable. Aquellas con $|t-valor|>2$ 
se consideran con un efecto significativo sobre la variable dependiente (ver \cite{bianchi2020human} para más detalles).

Partiremos de un modelo base (M0) que presenta variables típicamente utilizadas en el estudios de los movimientos oculares \cite{bianchi2020human,kliegl2004length}. Este modelo nos servirá para comprender los efectos de las variables de predictibilidad humana y computacional (ver \cite{bianchi2020human,umfurer2021using} para una discusión de los efectos de estas variables). 

Finalmente, cada uno de los modelos que incluyen comp-Pred será reanalizado con la cloze-Pred. Para esto, luego de ajustar el modelo obtendremos los residuos del mismo, al remover los efectos fijos estimados. Éstos residuos serán utilizados en un nuevo modelo lineal mixto con la Cloze-Pred como única variable.

\section{Resultado}
En la Tabla \ref{tabla} se presentan los resultados de los Modelos Lineales Mixtos realizados para diferentes combinaciones de variables. Los modelos M0 a M6 representan modelos analizados en trabajos previos. El M0 es el considerado Modelo Base, que cuenta con todas las variables no relacionadas con la Predictibilidad. El M1 cuenta con la cloze-Pred, es decir, es el modelo al que nos queremos acercar con las comp-Pred. El M2 tiene al modelo Ngram, que es hasta el momento el que mejor ha logrado modelar la Predictibilidad-Cloze. El M3, por su parte, es el mejor resultado obtenido en \cite{umfurer2021using} al entrenar (con Wikipedia) y reentrenar (con un gran corpus de textos narrativos) una red de tipo AWD-LSTM \cite{merityRegOpt2014}.

\begin{table}[h!]
\centering
\caption{\textbf{Resultados de los Modelos Lineales Mixtos:} cada columna representa un modelo lineal diferente. Todos los modelos fueron realizados sobre el mismo conjunto de datos. Los resultados presentados indican el t-valor obtenido para cada co-variable en cada uno de los modelos. La fila Cloze-Remef hace referencia al t-valor obtenido para el co-variable cloze-Pred al ser utilizada como único regresor en un LMM realizado sobre los residuos del modelo original.}
\label{tabla}
\resizebox{0.9\textwidth}{!}
{
\begin{tabular}{lccccccc}
\textbf{Co-variable} & \textbf{M0} & \textbf{M1} & \textbf{M2} & \textbf{M3} & \textbf{M4} & \textbf{M5} & \textbf{M6}\\
\textbf{} & Base & Cloze & Ngram & AWD epubs & GPT2 & GPT2 cuentos & GPT2 blogs\\
\hline
Dist. Sacada & 44.44 & 44.63 & 43.92 & 44.31 & 44.03 & 44.01 & 44.00\\
Longitud (inv) & -18.15 & -18.56 & -19.10 & -18.59 & -20.04 & -19.90 & -19.90\\
Frecuencia (log) & -10.83 & -10.60 & -1.93 & -5.77 & -4.98 & -5.12 & -5.01\\
Rel pos line & 4.14 & 3.96 & 4.33 & 4.12 & 4.45 & 4.56 & 4.61\\
Rel pos text & -3.93 & -3.28 & -3.87 & -4.68 & -4.56 & -4.15 & -4.25\\
Rel pos sntc & -5.36 & -4.85 & -5.76 & -4.97 & -3.83 & -3.81 & -3.92\\
Long:Frec & 16.98 & 17.17 & 15.68 & 16.91 & 15.59 & 15.58 & 15.61\\[7pt] 
Pred (logit) &  & -16.23 & -21.02 & -18.23 & -22.51 & -22.52 & -22.54\\[7pt] 
Cloze-Remef & -16.14 & 0.00 & -9.47 & -7.37 & -6.18 & -6.39 & -6.58
\end{tabular}
}
\end{table}

Los modelos correspondientes a las comp-Pred generadas por los modelos GPT2 (modelo original: M4, modelo reentrenado con cuentos: M5, modelo reentrenado con blogs argentinos: M6) muestran resultados muy similares entre ellos, y prometedores. En los tres modelos podemos ver que el t-valor obtenido es del orden del t-valor para la comp-Pred de Ngram (M2). Sin embargo, al analizar el impacto de estas comp-Preds sobre el resto de las co-variables (comparando con M0) podemos observar que los mismos sufren menos variaciones que en los M2 y M3. En particular, esto se observa en el efecto de la frecuencia léxica. Mientras que en M2 ésta pierde significancia, en los modelos M4, M5 y M6 se observan t-valores por encima del umbral. En este sentido, los modelos reentrenados parecen ser levemente mejor que el modelo original.

Por su parte, en lo que respecta al análisis de la varianza residual, podemos ver que la cloze-Pred todavía es capaz de capturar varianza suficiente para tener un efecto significativo. En este sentido, el modelos que mejor se desempaña en capturar esta varianza es el GPT2 original.

\section{Discusión}
El avance producido en los últimos años en los modelos del área del Procesamiento de Lenguaje (PLN) ha permitido que los mismos sean utilizados por la Neurociencia Cognitiva para profundizar el conocimiento de los procesos subyacentes del lenguaje \cite{toneva2019interpreting,wehbe2014aligning,huth2016natural,antonello2021low}. En trabajos anteriores realizamos avances en comprender cómo estos tipo de modelos, en particular los modelos de lenguaje causales, son capaces de modelar la cloze-Pred como co-variables en modelos estadísticos utilizados para comprender los movimientos oculares \cite{bianchi2020human,umfurer2021using}. En el presente trabajo profundizamos esta línea de investigación mediante el análisis de los resultados de la arquitectura GPT2 \cite{radford2018improving}.

Los resultados obtenidos en el presente trabajo muestran que el modelo GPT2 entrega las mejores comp-Preds hasta el momento. Este resultado fue independiente de si el modelo fue utilizado tal cual se encuentra en el repositorio de HuggingFace o si el mismo fue reentrenado con textos del dominio específico o de la variante del español de los lectores. Es importante recalcar dos puntos con respecto al reentrenamiento. Por un lado, el entrenamiento original del modelo GPT2 utilizado incluyó textos narrativos, por lo que el reentrenamiento puede no haber agregado información novedosa al modelo. Por otro lado, el reentrenamiento con el corpus en español rioplatense se realizó con un corpus muy chico (28MB) con respecto al corpus original (11GB). 

\bb{La mejoría de las comp-Pred basadas en GPT2 con respecto a los modelos utilizados previamente (en particular con AWD-LSTM) muestra los beneficios del uso  de \textit{transformers}, que parecería permitir capturar mayor información lingüística. En este sentido, los \textit{transformers} tienen la ventaja de poder analizar todo el texto de entrada a la vez, sin perder información de palabras lejanas. Además, el aumento en la complejidad de estos modelos en los últimos años también ha permitido que la comprensión que logran estas redes sea aún mayor. Esto permitido, principalmente, por un aumento sostenido en la cantidad de parámetros internos de los modelos. Mientras que nuestro modelo basado en AWD-LSTM cuenta con unos $10^7$ parámetros, GPT2 cuenta con $10^9$, es decir dos órdenes de magnitud de diferencia. 
}

En el futuro cercano nos proponemos aumentar el corpus correspondiente a español rioplatense, con el objetivo de profundizar el análisis de este tipo de reentrenamiento. Además, también planificamos profundizar y diversificar los análisis correspondientes a utilizar los resultados de éste tipo de modelos para mejorar la comprensión de los procesos mentales.

\section{Referencias}

\bibliographystyle{unsrt}
\renewcommand{\section}[2]{}%
{\footnotesize
\bibliography{biblio.bib}}








\end{document}